\documentclass[mathleft
]{an}
\usepackage{graphicx}
\usepackage{ulem,lscape,longtable,subfigure}
\usepackage{times}
\usepackage{txfonts}
\usepackage{natbib}
\bibpunct{(}{)}{;}{a}{}{,}
\def \deg {^\circ}
\def \corot {CoRoT}

\def \exodat {\textit{ExoDat}}
\def \aap {A\&A}

\def \na {NewA}
\def \icarus {Icarus}
\def \rmxaa {RMxAA}

\begin{document}

\Pagespan{789}{}
\Yearpublication{2006}%
\Yearsubmission{2005}%
\Month{11}%
\Volume{999}%
\Issue{88}%

\title{The power of low-resolution spectroscopy: On the spectral classification of planet candidates in the ground-based {\corot} follow-up}

\author{M. Ammler-von Eiff\inst{1,2}\fnmsep\thanks{Corresponding author:
  \email{ammler@mps.mpg.de}\newline}
\and D. Sebastian\inst{1}
\and E.W. Guenther\inst{1,3}
\and B. Stecklum\inst{1}
\and J. Cabrera\inst{4}
}
\titlerunning{Classification of {\corot} candidates}
\authorrunning{Ammler-von Eiff et al.}
\institute{Th\"uringer Landessternwarte, Sternwarte 5, 07778 Tautenburg, Germany              
             \and
             Max-Planck-Institut f\"ur Sonnensystemforschung, Justus-von-Liebig-Weg 3, 37077 G\"ottingen, Germany
             \and
\"Osterreichische Akademie der Wissenschaften, Institut f\"ur Weltraumforschung, IWF, Schmiedlstra{\ss}e 6, 8042 Graz, Austria
\and
Institute of Planetary Research, German Aerospace Center, Rutherfordstrasse 2, D-12489 Berlin, Germany}

\received{30 May 2005}
\accepted{11 Nov 2005}
\publonline{later}


\abstract{%
Planetary transits detected by the {\corot} mission can be mimicked by a low-mass star in orbit around a giant star. Spectral classification helps to identify the giant stars and also early-type stars which are often excluded from further follow-up.
  We study the potential and the limitations of low-resolution spectroscopy to improve the photometric spectral types of {\corot} candidates. In particular, we want to study the influence of the signal-to-noise ratio (SNR) of the target spectrum in a quantitative way. We built an own template library and investigate whether a template library from the literature is able to reproduce the classifications. Including previous photometric estimates, we show how the additional spectroscopic information improves the constraints on spectral type.
  Low-resolution spectroscopy ($R\approx$1000) of 42 {\corot} targets covering a wide range in SNR (1-437) and of 149 templates was obtained in 2012-2013 with the Nasmyth spectrograph at the Tautenburg 2m telescope. Spectral types have been derived automatically by comparing with 
the observed template spectra. The classification has been repeated with the external CFLIB library.
  The spectral class obtained with the external library agrees within a few sub-classes when the target spectrum has a SNR of about 100 at least. While the photometric spectral type can deviate by an entire spectral class, the photometric luminosity classification is as close as a spectroscopic classification with the external library. A low SNR of the target spectrum limits the attainable accuracy of classification more strongly than the use of external templates or photometry. Furthermore we found that low-resolution reconnaissance spectroscopy ensures that good planet candidates are kept that would otherwise be discarded based on photometric spectral type alone.}

\maketitle

\section{Introduction}
\label{sect:intro}

The {\corot}\footnote{\underline{Co}nvection, \underline{Ro}tation \& planetary \underline{T}ransits} mission has been the first space mission dedicated to the search for transiting planets \citep{2007AIPC..895..201B}. Overviews of the mission were given by \citet{Baglin+2009}, \citet{Deleuil+2011}, and \citet{Moutou+2013}. Statistics on candidates and planet detections have been reported for several fields monitored by {\corot} \citep{Carpano+2009,Moutou+2009a,Carone+2012,Erikson+2012,Cabrera+2009,Cavarroc+2012}.

The goals of ground-based follow-up are manifold \citep{Carpano+2009,Carone+2012}, among others the precise derivation of the parameters of a detected planet. While the {\corot} light curve constrains the radius of a planet, ground-based radial-velocity (RV) measurements are needed to derive its mass \citep{Moutou+2009a}. Only the relative radius and mass of the planet are constrained. Therefore, the absolute mass and the radius of the host star have to be known very accurately.

False positives play an important role and low-resolution spectroscopy is a step towards their identification. In the case of a false positive, a planetary transit is mimicked by other configurations \citep{Brown2003}. The case of a background eclipsing binary close to the actual target can be resolved by photometry \citep{Almenara+2009,Guenther+2013}. The present work aims at another kind of false positive. A planetary transit can be mimicked by a low-mass star orbiting a giant star \citep[see][]{Cavarroc+2012}. The identification of giant stars relies on the availability of accurate spectral types.

A first clue about the spectral type of {\corot} targets is given by the {\corot} input catalogue (\exodat\footnote{http://cesam.oamp.fr/exodat/}; \citealp{Deleuil+2009}) which is based on a massive $UVBr'i'$ photometric survey carried out during the mission preparatory phase and has been updated continuously since then. However, in regions of inhomogeneous extinction, it is difficult to obtain spectral class, luminosity class, and extinction simultaneously. Although the photometric classification is correct on average, it can be deviant for individual stars \citep{sebastian12}. Therefore, spectroscopic information is needed to better assess stellar parameters of individual targets \citep[cf.][]{Carone+2012,Gazzano+2010a}.

Precise information is contained in atmospheric parameters like effective temperature and surface gravity. Those are obtained most accurately from high-resolution spectroscopy \citep[e.g.][for {\corot}]{Gazzano+2013a}. As the necessary signal-to-noise ratio (SNR) can be hardly attained at high spectral resoluton for faint stars, one needs to resort to low-resolution spectroscopy. Then, a precise measurement of atmospheric parameters is out of reach but spectral classification is still possible.

In practice, spectral classification is done by comparing stellar spectra to template spectra of well-known stars, in particular MK standards. There are different approaches of how to obtain template spectra and to compare them to the target spectra. Computer-based classification has proven efficient. \citet{sebastian12} classified more than 10,000 stars in the fields of CoRoT automatically and showed that an accuracy of one to two sub-classes can be achieved. While templates are commonly taken from libraries taken with other instruments \citep[e.g.][]{2008ApJ...687.1303G,sebastian12}, the classification is considered most accurate when the templates are taken from an internal library, i.e. have been observed with the same instrument and setup as the target spectra \citep{2009ssc..book.....G,2011RAA....11..924W}. Then, the full spectral range is available for comparison and no convolution is needed  to match the spectral resolution.

A low-resolution spectrograph attached to an intermediate-size telescope, like the Nasmyth spectrograph ($R\approx1,000$) at the Tautenburg 2m telescope, offers a sufficient dynamical range to observe the faint {\corot} targets ($V<16\,$mag) as well as bright nearby template stars. The Nasmyth spectrograph provides a wide spectral range $360-935\,$nm covering a wealth of stellar features which can be used for classification.


In the present study, we follow the initial work of \citet{Guenther+2012} and \citet{sebastian12}. Based on spectra taken with the Nasmyth spectrograph, we explore the capabilities of low-resolution spectroscopy to derive the spectral types of {\corot} targets. Of course, the accuracy of spectral classification will depend on the signal-to-noise ratio (SNR) of the target spectra. In the course of the follow-up of planet candidates, a large number of stars has to be classified efficiently and to the accuracy required. Since telescope time is limited and the exposure time scales with the square of the SNR, we are interested in a good knowledge of the required SNR.

The {\corot} targets were selected from various internal and public lists, in particular from Deleuil et al. (in preparation) \footnote{The present work includes a statistical study on the type and quality of information available from spectral classification. Therefore, additional follow-up information on particular planet candidates is not included here and a discussion of the status of individual candidates is beyond the scope of the present work (but see Deleuil et al.).}. Most of the {\corot} objects are FGK stars \citep{Guenther+2012} and these are the most interesting planet host stars. How well does spectral classification distinguish giant stars and dwarf stars, especially at spectral types F, G, and K?

We followed a two-fold approach. Firstly, we followed the canonical approach of spectral classification \citep{2009ssc..book.....G} and built an internal library of spectral templates with the same spectrograph and setup used to observe the {\corot} targets. Secondly, a classification was done using templates from the Indo-US Coud\'e-feed library \citep[CFLIB,][]{valdes04}, selected for its good coverage of spectral types, luminosity classes, and metallicity. We investigate whether this external library gives similar results. Simultaneously, we analyse the impact of the SNR of the target spectra on the results. In addition, we compare the spectral types obtained to previous photometric classifications given by {\exodat}. We identify the merits of the inclusion of a spectroscopic classification over a sole photometric classification.

Section~\ref{sect:temp} explains the selection of stars to build the internal template library. The observation of templates and {\corot} targets is described in Sect.~\ref{sect:obs}. Sect.~\ref{sect:red} addresses the data reduction and the coverage of the internal template catalogue is assessed in Sect.~\ref{sect:coverage}. We describe the steps of the spectral classification in Sect.~\ref{sect:class} and the results in Sect.~\ref{sect:results}. The results are discussed in Sect.~\ref{sect:disc} before we conclude in Sect.~\ref{sect:concl}.

\section{Selecting stars for a new library of template spectra}
\label{sect:temp}

The template stars need to fulfil a set of requirements. Of course, the spectral types need to be known very well. Having in mind quantitative spectral classification in future work, we adopted as an additional criterion the availability of accurate atmospheric parameters. 

Although quantitative classification is beyond of the scope of the present work, we point out here that it has many advantages over the use of spectral types. There is no need to use a conversion scale, e.g. from spectral type to effective temperature which is intrinsically prone to errors. Furthermore, it promises higher precision and flexibility than the classification by spectral type which is limited by the discreteness of the classification scheme. An extension of the traditional grid of spectral types to include metallicity or even abundance patterns would be too demanding. Instead, atmospheric parameters can be used defining a continuous grid which can be sampled by templates according to the requirements of the classification task. For previous work and reviews, see \citet{Cayrel+1991,Gray+1991,Stock+1999,Malyuto+2001,Bailer-Jones2002, Singh+2002}.

As another criterion, the template stars should be bright so that spectra can be taken quickly with a high SNR. The templates should cover the range of spectral types of interest. 
In order to be able to classify most {\corot} targets, we selected templates for early-type and Sun-like stars of different luminosity class from the CFLIB. \citet{Wu+2011a} obtained stellar parameters and chemical abundance homogeneously for the CFLIB library. 

FGK dwarfs are the best targets to look for planets with {\corot}. To fill the grid with templates, we selected well-studied FGK stars from Fuhrmann (1998-2011)\nocite{Fuhrmann1998,Fuhrmann2000,Fuhrmann2004,Fuhrmann2008,Fuhrmann2011}. Several of those are MK standards according to \citet{2009ssc..book.....G}.

The work of Fuhrmann is restricted to solar-type main-sequence and sub-giant stars. However, in the distant {\corot} fields, the fraction of early-type stars and giants is relatively high. {\corot} covers these luminous stars more completely but there is a lack of cooler main-sequence stars. While early-type stars are covered by the CFLIB, we supplement the set of stellar templates with K giants compiled by \citet{Doellinger2008} who derived the stellar parameters and iron abundance of 62 K giants.

\section{Observations}
\label{sect:obs}

\begin{table}
\caption{\label{tab:specinfo} Journal of observations sorted by the SNR achieved. The first two columns give the CoRoT and Win identifiers. The Win identifier is composed of the CoRoT run (LR=long run, SR=short run, IR=initial run; a=galactic anti-centre, c=galactic centre; plus a consecutive number), the CCD identifier (E1/E2), and a consecutive number. The run ID identifies the spectroscopic runs carried out in February and July 2012/2013. The $R$ band magnitude is followed by the number of spectra taken per object. The last column gives the SNR of the co-added spectra.}
\begin{tabular}{rllrrr}
\hline
{\corot} ID&Win ID&run ID &$R$&\#&SNR\\
\hline
102634864&LRa01\_E1\_3221&Feb12& 15.2& 3&  1\\                                                                                                                                                                              
102627709&LRa01\_E2\_1578&Feb12& 13.3& 1&  8\\                                                                                                                                                                              
102580137&LRa01\_E2\_4519&Feb12& 15.4& 3& 16\\                                                                                                                                                                              
102698887&LRa01\_E1\_2240&Feb12& 14.2& 3& 22\\                                                                                                                                                                              
616759073&SRa05\_E2\_3522&Feb12& 13.9& 3& 37\\                                                                                                                                                                              
310153107&LRc03\_E2\_5079&Jul13& 14.0& 1& 40\\                                                                                                                                                                              
221699621&SRa02\_E1\_1011&Feb12& 14.0& 2& 60\\                                                                                                                                                                              
310155742&LRc03\_E2\_5451&Jul13& 14.6& 3& 71\\                                                                                                                                                                              
678658629&LRc09\_E2\_3403&Jul13& 14.7& 3& 71\\                                                                                                                                                                              
655228061&LRc09\_E2\_2479&Jul13& 14.5& 2& 73\\                                                                                                                                                                              
631424752&LRc07\_E2\_2968&Jul13& 13.7& 2& 78\\                                                                                                                                                                              
659668516&LRc08\_E2\_4203&Jul12& 14.6& 2& 80\\                                                                                                                                                                              
655049038&LRc09\_E2\_0892&Jul13& 13.9& 1& 82\\                                                                                                                                                                              
659714254&LRc07\_E2\_4203&Jul13& 14.9& 3& 83\\                                                                                                                                                                              
102901032&LRa02\_E1\_4967&Feb13& 15.5& 4& 89\\                                                                                                                                                                              
679896622&LRc09\_E2\_3338&Jul13& 15.2& 4& 89\\                                                                                                                                                                              
632089337&LRc10\_E2\_3956&Jul13& 13.9& 2&101\\                                                                                                                                                                              
738282899&LRa07\_E2\_3354&Feb13& 15.3& 4&103\\                                                                                                                                                                              
659676254&LRc08\_E2\_4520&Jul13& 14.6& 3&105\\                                                                                                                                                                              
102770212&LRa06\_E2\_5287&Feb13& 15.5& 4&107\\                                                                                                                                                                              
103970832&LRc04\_E2\_5713&Jul13& 15.5& 6&109\\                                                                                                                                                                              
315219144&SRa03\_E2\_1073&Feb13& 14.5& 3&125\\                                                                                                                                                                              
104833752&LRc05\_E2\_3718&Jul12& 14.7& 5&126\\                                                                                                                                                                              
680074530&LRc09\_E2\_0548&Jul12& 13.6& 2&126\\                                                                                                                                                                              
310170040&LRc03\_E2\_0935&Jul13& 13.8& 3&130\\                                                                                                                                                                              
633100731&LRc10\_E2\_5093&Jul13& 13.8& 4&132\\                                                                                                                                                                              
659718186&LRc10\_E2\_0740&Jul13& 13.1& 2&134\\                                                                                                                                                                              
633496006&LRc10\_E2\_1984&Jul13& 14.5& 1&141\\                                                                                                                                                                              
659473289&LRc09\_E2\_0308&Jul12& 11.0& 1&141\\                                                                                                                                                                              
631423929&LRc07\_E2\_0158&Jul12& 12.2& 2&179\\                                                                                                                                                                              
652180991&LRc08\_E2\_0275&Jul12& 13.3& 4&179\\                                                                                                                                                                              
 21160782&SRc01\_E1\_0346&Jul13& 12.7& 1&187\\                                                                                                                                                                              
659714295&LRc07\_E2\_0534&Jul12& 12.4& 2&196\\                                                                                                                                                                              
678656772&LRc09\_E2\_0131&Jul12& 12.3& 2&219\\                                                                                                                                                                              
652345526&LRc07\_E2\_0307&Jul12& 13.0& 4&226\\                                                                                                                                                                              
632279463&LRc07\_E2\_0146&Jul12& 12.6& 2&240\\                                                                                                                                                                              
652312572&LRc07\_E2\_0182&Jul12& 12.0& 2&248\\                                                                                                                                                                              
105314448&LRc06\_E2\_0119&Jul12& 12.8& 4&265\\                                                                                                                                                                              
631900113&LRc07\_E2\_0482&Jul12& 12.8& 4&277\\                                                                                                                                                                              
631423419&LRc07\_E2\_0187&Jul13& 12.3& 2&293\\                                                                                                                                                                              
104992379&LRc05\_E2\_0168&Jul12& 12.0& 2&322\\                                                                                                                                                                              
104992379&LRc05\_E2\_0168&Jul13& 12.0& 2&437\\                                                                                                                                                                              

\hline
\end{tabular}
\end{table}

\begin{table*}
\caption{\label{tab:templates} Stellar templates observed with the Nasmyth spectrograph and their spectral type. The second column gives the run ID which identifies the spectroscopic runs carried out in February and July 2012/2013.}
\begin{tabular}{rll|rll|rll}
\hline
HD number & run ID & spectral type&HD number & run ID & spectral type&HD number & run ID & spectral type\\
&&&&&&&&\\
\hline
   400&Jul12&F8IV           &      94028&Feb12&F4V            &     168151&Jul13&F5V            \\
  2628&Feb12&A7III          &      94084&Feb12&K2III          &     168723&Jul13&K0III-IV       \\
  6397&Jul12&F4II-III       &      95128&Feb12&G1V            &     170693&Jul13&K1.5III        \\
  6920&Jul12&F8V            &      96064&Feb12&G8V            &     172340&Jul13&K4III          \\
 10362&Feb12&B7II           &      96833&Jul13&K1III          &     173667&Feb12&F6V            \\
 10476&Jul12&K1V            &      97989&Feb12&K0III          &     173936&Feb12&B6V            \\
 10697&Jul12&G5IV           &      98281&Feb12&G8V            &     175823&Jul13&K5III          \\
 12303&Feb12&B8III          &      99491&Feb12&K0IV           &     176377&Jul13&G0             \\
 12846&Jul12&G2V            &     100563&Feb12&F5V            &     176524&Jul13&K0III          \\
 12953&Feb12&A1Iae          &     101501&Feb12&G8V            &     178187&Feb12&A4III          \\
 15318&Feb12&B9III          &     102328&Feb12&K3III          &     180554&Feb12&B4IV           \\
 30614&Feb12&O9.5Iae        &     102870&Feb12&F9V            &     180610&Jul13&K2III          \\
 34578&Feb12&A5II           &     103095&Feb12&G8Vp           &     182488&Jul13&G8V            \\
 35296&Feb12&F8V            &     104985&Feb12&G9III          &     182572&Jul13&G8IV           \\
 37394&Feb12&K1V            &     105546&Feb12&G2IIIw         &     183144&Feb12&B4III          \\
 38114&Feb12&G5             &     105631&Feb12&K0V            &     184385&Jul13&G8V            \\
 39866&Feb12&A2II           &     110897&Jul12&G0V            &     184499&Jul12&G0V            \\
 41117&Feb12&B2Iaevar       &     117043&Feb12&G6V            &     185144&Jul13&G9V            \\
 41692&Feb12&B5IV           &     117176&Feb12&G5V            &     185269&Jul12&G0IV           \\
 43247&Feb12&B9II-III       &     120136&Jul12&F6IV           &     186307&Feb12&A6V            \\
 43384&Feb12&B3Ib           &     121560&Feb12&F6V            &     186408&Jul13&G1.5V          \\
 47839&Feb12&O7Ve           &     122408&Feb12&A3V            &     186427&Jul13&G3V            \\
 47914&Jul13&K5III          &     127334&Feb12&G5V            &     186815&Jul13&K2III          \\
 48682&Feb12&G0V            &     128167&Feb12&F2V            &     187013&Jul12&F7V            \\
 50420&Feb12&A9III          &     130948&Feb12&G1V            &     187691&Jul13&F8V            \\
 51530&Feb12&F7V            &     132142&Jul12&F7V            &     187923&Jul13&G0V            \\
 52711&Feb12&G4V            &     132254&Feb12&F7V            &     187961&Jul12&B7V            \\
 55575&Feb12&G0V            &     136202&Feb12&F8III-IV       &     188510&Jul12&G5Vw           \\
 58855&Feb12&F6V            &     136729&Feb12&A4V            &     188512&Jul12&G9.5IV         \\
 59747&Feb12&G5V            &     139323&Feb12&K3V            &     189944&Feb12&B4V            \\
 59881&Feb12&F0III          &     142373&Jul12&F8Ve           &     190228&Jul12&G5IV           \\
 61295&Feb12&F6II           &     151862&Feb12&A1V            &     191243&Feb12&B5Ib           \\
 62301&Feb12&F8V            &     152614&Feb12&B8V            &     192699&Jul12&G5             \\
 63433&Feb12&G5IV           &     153653&Feb12&A7V            &     193322&Feb12&O9V            \\
 65583&Feb12&G8V            &     154431&Feb12&A5V            &     195810&Feb12&B6III          \\
 67228&Feb12&G1IV           &     154445&Feb12&B1V            &     196504&Feb12&B9V            \\
 72946&Feb12&G5V            &     155514&Feb12&A8V            &     205139&Feb12&B1II           \\
 74280&Feb12&B3V            &     157214&Jul12&G0V            &     206827&Jul12&G2V            \\
 75732&Feb12&G8V            &     157681&Jul13&K5III          &     207970&Jul12&F6IV\_Vwvar    \\
 76151&Feb12&G3V            &     158148&Feb12&B5V            &     208947&Feb12&B2V            \\
 82443&Feb12&K0V            &     158633&Jul12&K0V            &     210839&Feb12&O6Iab          \\
 82621&Feb12&A2V            &     160290&Jul13&K1III          &     210855&Jul12&F8V            \\
 84737&Feb12&G0.5Va         &     161797&Jul13&G5IV           &     212978&Feb12&B2V            \\
 85235&Feb12&A3IV           &     162570&Feb12&A9V            &     215648&Jul12&F7V            \\
 86728&Feb12&G3Va           &     164136&Feb12&F2II           &     216385&Jul12&F7IV           \\
 88983&Feb12&A8III          &     164922&Jul12&K0V            &     218470&Jul12&F5V            \\
 89269&Feb12&G5             &     165401&Jul13&G0V            &     221830&Jul12&F9V            \\
 89744&Feb12&F7V            &     167042&Jul13&K1III          &     222794&Jul12&G2V            \\
 90277&Feb12&F0V            &     168009&Jul13&G2V            &     223385&Feb12&A3Iae          \\
 91752&Feb12&F3V            &     168092&Feb12&F1V            &     &&\\                             

\hline
\end{tabular}
\end{table*}

All spectra were taken with the low-resolution long-slit spectrograph mounted at the Nasmyth focus of the 2m Alfred Jensch telescope in Tautenburg (Germany). A slit width of $1''$ was used to ensure a resolution of $\approx1,000$. The V200 grism was chosen as dispersing element, thus covering the visual wavelength range from 360 to 935\,nm. It is a BK7 grism with 300 lines per millimeter and a dispersion of $225\,\mathrm{\AA}\mathrm{mm}^{-1}$. The detector is a SITe\#T4 CCD with 2048x800 pixels and a pixel size of $15\,\mathrm{{\mu}m}$. We used channel A with a gain of $1.11$\,$\mu$V per electron. 

{\corot} monitors several stellar fields consecutively for up to 40\,days (short runs) or for up to 150\,days (long runs). These fields are selected in two opposite directions in the sky where the galactic plane crosses the equatorial plane. These are the so-called galactic “centre” and “anti-centre”-eyes of {\corot}. Seasonal observations of the {\corot} fields are possible in winter (galactic anti-centre) and summer (galactic centre) despite the high latitude of the observing site of $51\deg$N. The {\corot} fields can be observed at an air mass of approx. 1.5. A total of 42 {\corot} targets covering a brightness range of $R=11.0-15.5$ was observed in February and July 2012/2013 when the visibility of the {\corot} eyes was best. Table~\ref{tab:specinfo} presents the journal of observations.

In addition, we have observed a set of 149 template stars (Table~\ref{tab:templates}). They are bright so that the SNR is high. Exposure times range from a few seconds for the brightest template stars to 30\,min for the faintest {\corot} targets. In general, at least two spectra were taken per target in order to remove cosmics. While the usually faint {\corot} targets are observable only under good conditions for a few hours per night, the bright template stars are distributed over the whole sky. They perfectly fill the observing time when {\corot} targets are not visible or when conditions are bad.

In order to estimate the bias level and the read-out noise, dark frames were taken with closed shutter and zero exposure time. In order to assess the pixel-to-pixel variation of sensitivity, the wavelength-dependent transmission of the spectrograph, and bad or hot pixels, flat-field frames were taken pointing the telescope to a flat-field screen which is installed in the dome and illumated by an incandescent lamp.

In principle, the wavelength calibration can be done using sky emission lines in the long-slit spectra. However, the sparseness of sky emission lines in the blue part of the spectra does not permit a precise calibration in this region. It is the blue part of the spectrum which displays a wealth of features and is very valuable for spectral classification. Therefore, spectra of gas discharge lamps (He and Kr) were used for wavelength calibration. At least one set of gas discharge exposures was taken per observing run. A wavelength precision of a few {\AA} is achieved this way and is sufficient for the purpose of spectral classification.

To estimate the signal-to-noise ratio (SNR), we assumed pure photon noise accounting for read-out noise. The total system gain used depends on the amplifier setting chosen. In the present case (setting '20'), we estimated a total system gain of 3.2 electrons per data unit using the full well at the noise roll-over specified by the manufacturer (84,000 electrons)\footnote{The full well was scaled to the saturation level obtained at setting '50' and then rescaled to setting '20'.}. The readout-noise amounts to 7.8 electrons. This estimate of the SNR is fully sufficient to compare the spectra since all spectra were taken with the same instrument and identical settings. The median is used to distinguish good spectra with high signal and bad spectra with a high noise level.

\section{Data reduction}
\label{sect:red}

The long-slit spectra have been reduced using the data reduction and analysis system IRAF\footnote{IRAF is distributed by the National Optical Astronomy Observatories, which are operated by the Association of Universities for Research in Astronomy, Inc., under cooperative agreement with the National Science Foundation.} \citep{iraf86,Tody1993}. This procedure comprises bias subtraction, an automatic removal of cosmic rays, and removal of pixel-to-pixel variations by flat-field correction.

All spectra have been wavelength-calibrated using He and Kr spectra. Using IRAF tasks, the wavelengths of prominent lines have been identified and used to derive a wavelength solution which has been applied to all stellar spectra. The spectrograph is mounted at the fork of the telescope so that instrument flexure causes shifts depending on the exact orientation of the telescope. Therefore, sky emission lines have been employed to apply corrections. For each frame with a long exposure time of 20 minutes at least, the sky lines are bright enough to use them as calibration source \citep{osterbrock96}. They have been used to derive offsets to correct the wavelength scale. The corrections were small and of the order of the resolution limit.

\begin{figure}[h] 
\centering
\includegraphics[width=8cm]{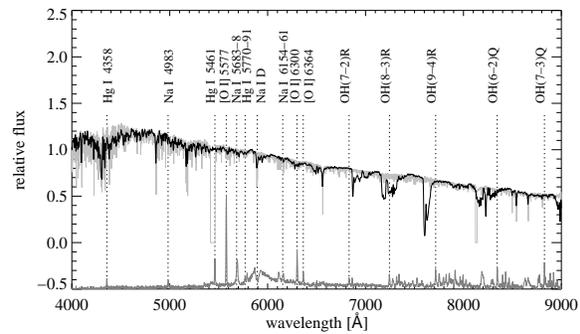}
\caption{The black line shows a Nasmyth spectrum of the G0V star \object{HD\,157214}. For comparison, a library spectrum with medium resolution \citep{valdes04} is shown (light grey). To match the SED, the template spectrum was used to adjust the flux of the Nasmyth spectrum. Both spectra were normalised to unity at $5550\,${\AA}. In addition, the subtracted night sky spectrum of the site is shown (grey) and is offset w.r.t. the stellar spectra for clarity. The most important spectral features are indicated.}
\label{fig:sky}
\end{figure}

The sky background could be subtracted easily from the long-slit spectra. For this purpose, long exposures provide an excellent source to analyse the night-sky emission in Tautenburg. Figure~\ref{fig:sky} shows the Nasmyth spectrum of a Sun-like star together with the subtracted sky-background.

The most prominent features in the night-sky spectrum are the O\,I, Hg\,I, and Na\,I atomic lines, the broad continuum centred at 5890\AA, and the OH bands in the red part \citep{Osterbrock+1992,Slanger+2003}.
These lines originate from the night glow as well as from artificial sources. 
The broad continuum centered at 5890\,{\AA} is characteristic of the presence of artificial light and in this case originates from high-pressure sodium lamps used in the nearby city of Jena, Germany.

The flux calibration of the spectra of {\corot} targets was not possible because of the spectrograph design and the high airmass of {\corot} targets. The orientation of the spectrograph slit cannot be adjusted and the slit-rotation of the spectrograph depends on the hour angle of the object. This leads to a dependency of the detected flux level on atmospheric refraction. In addition, the star is slightly moving on the slit and the exact location on the slit cannot be reproduced. Therefore, the stellar continuum could not be recovered and used for classification. Instead, one has to rely on spectral lines only and the continuum has to be adjusted to match the continuum of the template (Fig.~\ref{fig:spec_class}). Hence, the comparison is restricted to the strength and the profile of selected absorption lines.

\section{The coverage of the new template library}
\label{sect:coverage}

The CFLIB and the new internal template library have many stars in common. 54 CFLIB stars have been included in the sample chosen from Fuhrmann (1998-2011). The Fuhrmann sample does not cover all spectral types. Therefore, 55 stars have been chosen from the CFLIB catalogue and reobserved for the new library. In total, 109 CFLIB stars have been reobserved. 

Within the present work, 149 template spectra of bright stars have been taken. MK spectral types have been adopted from \citet{valdes04} when available and from the SIMBAD database otherwise.

Table~\ref{tab:grid_spt} shows the coverage in spectral type. The catalogue is being filled continuously but a good coverage has already been achieved for dwarf stars. Particularly, FGK dwarfs are densely covered. A finely graduated classification becomes feasible in the FGK regime.

\begin{table}
\caption{\label{tab:grid_spt} The coverage of the new template library in spectral type. At each spectral type, the number of templates is given.}
\begin{center}
\begin{tabular}{l|rrrrrrr}
\hline
spectral&\multicolumn{7}{c}{luminosity class}\\
class&I&II&II-III&III&III-IV&IV&V\\
\hline
O6& 1&  &  &  &  &  &  \\
O7&  &  &  &  &  &  & 1\\
O9&  &  &  &  &  &  & 1\\
O9.5& 1&  &  &  &  &  &  \\
\hline
B1&  & 1&  &  &  &  & 1\\
B2& 1&  &  &  &  &  & 2\\
B3& 1&  &  &  &  &  & 1\\
B4&  &  &  & 1&  & 1& 1\\
B5& 1&  &  &  &  & 1& 1\\
B6&  &  &  & 1&  &  & 1\\
B7&  & 1&  &  &  &  & 1\\
B8&  &  &  & 1&  &  & 1\\
B9&  &  & 1& 1&  &  & 1\\
\hline
A1& 1&  &  &  &  &  & 1\\
A2&  & 1&  &  &  &  & 1\\
A3& 1&  &  &  &  & 1& 1\\
A4&  &  &  & 1&  &  & 1\\
A5&  & 1&  &  &  &  & 1\\
A6&  &  &  &  &  &  & 1\\
A7&  &  &  & 1&  &  & 1\\
A8&  &  &  & 1&  &  & 1\\
A9&  &  &  & 1&  &  & 1\\
\hline
F0&  &  &  & 1&  &  & 1\\
F1&  &  &  &  &  &  & 1\\
F2&  & 1&  &  &  &  & 1\\
F3&  &  &  &  &  &  & 1\\
F4&  &  & 1&  &  &  & 1\\
F5&  &  &  &  &  &  & 3\\
F6&  & 1&  &  &  & 2& 3\\
F7&  &  &  &  &  & 1& 6\\
F8&  &  &  &  & 1& 1& 6\\
F9&  &  &  &  &  &  & 2\\
\hline
G0&  &  &  &  &  & 1& 8\\
G0.5&  &  &  &  &  &  & 1\\
G1&  &  &  &  &  & 1& 2\\
G1.5&  &  &  &  &  &  & 1\\
G2&  &  &  & 1&  &  & 4\\
G3&  &  &  &  &  &  & 3\\
G4&  &  &  &  &  &  & 1\\
G5&  &  &  &  &  & 4& 8\\
G6&  &  &  &  &  &  & 1\\
G8&  &  &  &  &  & 1& 8\\
G9&  &  &  & 1&  &  & 1\\
G9.5&  &  &  &  &  & 1&  \\
\hline
K0&  &  &  & 2& 1& 1& 4\\
K1&  &  &  & 3&  &  & 2\\
K1.5&  &  &  & 1&  &  &  \\
K2&  &  &  & 3&  &  &  \\
K3&  &  &  & 1&  &  & 1\\
K4&  &  &  & 1&  &  &  \\
K5&  &  &  & 3&  &  &  \\

\hline
\end{tabular}
\end{center}
\end{table}

\section{Computer-based classification}
\label{sect:class}

\begin{figure}[h]
\centering
\includegraphics[width=4cm,angle=270]{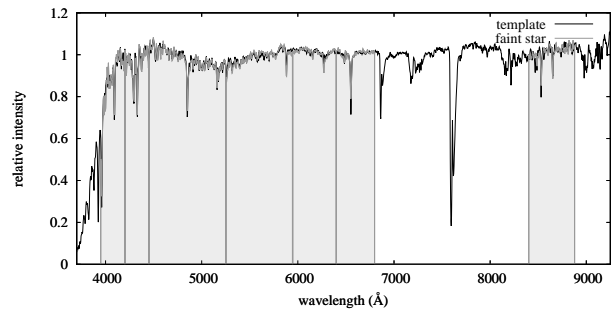}
\caption{Comparison of a target spectrum (grey line) with the 
best-matching template (black line). The vertical lines mark the borders of chunks which are compared separately and used to adjust the continuum level of the spectra.}
\label{fig:spec_class}
\end{figure}

The spectral types of the CoRoT targets have been obtained using a computer-based classification. 
The software applied (described in \citealp{sebastian12}) compares all target spectra to a 
library of template spectra. Each target spectrum is matched to template spectra
adjusting the radial velocity shift, the level of the continuum, and the slope of the continuum.  We selected spectral chunks that contain sensitive lines for spectral classification thus removing 
parts that are not useful for classification (see Fig.~
\ref{fig:spec_class}).
Afterwards the ${\chi}^2$ is calculated and compared for each template spectrum. 
The five best-matching templates are validated by visual inspection. This 
validation rules out false classifications due to low S/N or stellar 
activity. The spectral type of the best-matching validated template is adopted for the 
target spectrum.
\citet{Guenther+2012} found that the error bar of this classification 
method depends on the spectral type and the template library used. They analysed the 
accuracy of the method on a sample of more than 3000 stars and found that the accuracy of this method for low resolution spectra is 
on average two subclasses. It is slightly better for 
early-type stars (1.3 subclasses) and less accurate for solar-type 
stars (2-3 subclasses). The classification is not affected by rotational broadening since the spectral resolution of the Nasmyth spectrograph is too low.

Firstly, we used the internal template library taken
with the Nasmyth spectrograph. In the second approach, we
used a set of 281 template spectra provided by CFLIB 
(Valdes et al. 2004)\footnote{These CFLIB templates have a high SNR of $100$ at least and cover almost the full range of spectral types.}.
To compare the {\corot} target spectra directly with the CFLIB templates, we have to ensure that the resolution, the wavelength range, and the continuum flux level are roughly the same. Since the Nasmyth spectra have been taken with a resolution of $\sim5$\,{\AA} FWHM (at $5500$\,{\AA}), we convolved the CFLIB templates ($\sim1$\,{\AA} FWHM) with a Gaussian kernel.


For the comparison with internal templates, the full wavelength range from $3600$ to $9350\,${\AA} can be used in principle since the target spectra have been obtained with the same spectrograph. However, especially the red part of the covered wavelength region contains telluric bands. The strength of these bands is variable. Although a correction of telluric bands is possible in principle, the scarcity of useful features does not justify this effort. The strongest telluric oxygen bands appear at $6884\,${\AA} \citep{catanzaro97}. Therefore, the wavelength region applied here starts in the blue at 3950\,{\AA} where we get sufficient signal and ends at 6800\,{\AA} at the blue edge of the oxygen bands. In the extreme red, there is a very short range of $\approx400$\,{\AA} not strongly affected by telluric absorption. This region contains three prominent CaII lines ($8498$, $8542$, and $8662\,${\AA}) at almost all spectral types which can be used for spectral classification \citep{2009ssc..book.....G}. For early-type stars, some HI lines of the Paschen series were included in addition. In spite of some night-sky emission lines that could affect the spectra (see Fig.~\ref{fig:sky}), we chose the region from 8400 to 8880\,{\AA} in addition to the blue part ($4200-6800${\,\AA}) for comparison  (Fig.~\ref{fig:spec_class}). For spectra with low SNR, we had to set the blue edge to $4800\,${\AA} to skip the blue part with very low signal.

When using the external CFLIB templates, the wavelength range is limited to the overlapping region of the {\corot} target spectra and the CFLIB templates. In particular, the very red part with the Ca\,II triplet and the Paschen series cannot be used. Moreover, parts with strong telluric absorption had to be excluded again and once more, the wavelength range had to be reduced to $4800-6850$\,{\AA} in the case of noisy spectra.

The design of the spectrograph does not allow one to reproduce the continuum flux (Sect.~\ref{sect:red}) so that the continuum flux could not be used for classification. Therefore, the {\corot} spectra and the Nasmyth templates were normalised and seven chunks (five in the case of low SNR, resp.) were compared separately (Fig.~\ref{fig:spec_class}). This way, the influence of the unknown absolute flux level and the extinction was optimally removed. This method ensures that the spurious continuum flux levels of both spectra do not distort the results of the classification. 

The CFLIB templates are not flux-calibrated either but \citet{valdes04} recovered the SED of the CFLIB spectra from spectrophotometric templates. In order to compare the Nasmyth spectra and the CFLIB templates, they were normalised again in an appropriate way and compared in different chunks.

Figure~\ref{fig:spec_class} shows the comparison of a {\corot} target  to its best-matching Nasmyth template. In this case, the target spectrum matches the spectrum of HD 121560 perfectly (spectral type F6V).

\section{Results}
\label{sect:results}

\begin{table*}
\caption{\label{tab:classinfo} Results of spectral classification (sorted by SNR). The table contrasts the classification obtained via two different sets of templates -- the internal library obtained by us with the Nasmyth spectrograph and the template library of \citet{valdes04} (CFLIB). \object{LRc10\_E2\_1984} is an M star for which no template was taken with the Nasmyth spectrograph.}
\renewcommand{\tabcolsep}{0.3em}
\footnotesize
\centering
\begin{tabular}{lrl|ll|ll}
\hline
&&&\multicolumn{2}{c|}{templates taken with }&\multicolumn{2}{c}{templates taken from} \\
&&&\multicolumn{2}{c|}{the Nasmyth spectrograph}&\multicolumn{2}{c}{\citet{valdes04}}\\
Win ID&SNR&spec. type&templ.&spec.&templ.&spec.\\
&&(Exodat)&&type&&type\\
\hline
LRa01\_E1\_3221&  1&A5V       &HD139323  &K3V       &HD190390  &F1III     \\                        
LRa01\_E2\_1578&  8&A5V       &HD210855  &F8V       &HD158352  &A8V       \\                        
LRa01\_E2\_4519& 16&A5IV      &HD136729  &A4V       &HD173495  &A1V+      \\                        
LRa01\_E1\_2240& 22&A5V       &HD61295   &F6II      &HD150453  &F3V       \\                        
SRa05\_E2\_3522& 37&O5V       &HD136729  &A4V       &HD177724  &A0Vn      \\                        
LRc03\_E2\_5079& 40&G5II      &HD47914   &K5 III    &HD232078  &K3IIp     \\                        
SRa02\_E1\_1011& 60&F8IV      &HD102870  &F9V       &HD16673   &F6V       \\                        
LRc03\_E2\_5451& 71&K3I       &HD99491   &K0IV      &HD34255   &K4Iab:    \\                        
LRc09\_E2\_3403& 71&G8V       &HD84737   &G0.5Va    &HD177249  &G5.5IIb   \\                        
LRc09\_E2\_2479& 73&K0III     &HD47914   &K5 III    &HD178717  &K3.5III:  \\                        
LRc07\_E2\_2968& 78&G5III     &HD184385  &G8V       &HD5286    &K1 IV     \\                        
LRc08\_E2\_4203& 80&A0V       &HD2628    &A7III     &HD186377  &A5III     \\                        
LRc09\_E2\_0892& 82&K5III     &HD160290  &K1III     &HD175306  &G9 IIIb   \\                        
LRc07\_E2\_4203& 83&G0IV      &HD84737   &G0.5Va    &HD39833   &G0 III    \\                        
LRa02\_E1\_4967& 89&A5IV      &HD128167  &F2V       &HD204363  &F7V       \\                        
LRc09\_E2\_3338& 89&A0V       &HD136729  &A4V       &HD177724  &A0 Vn     \\                        
LRc10\_E2\_3956&101&K2III     &HD84737   &G0.5Va    &HD191615  &G8 IV     \\                        
LRa07\_E2\_3354&103&A2V       &HD39866   &A2II      &HD168270  &B9V       \\                        
LRc08\_E2\_4520&105&A5IV      &HD84737   &G0.5Va    &HD128987  &G6V       \\                        
LRa06\_E2\_5287&107&G0IV      &HD168151  &F5V       &HD150453  &F3V       \\                        
LRc04\_E2\_5713&109&A5IV      &HD165401  &G0V       &HD130948  &G1 V      \\                        
SRa03\_E2\_1073&125&F8V       &HD91752   &F3V       &HD56986   &F0IV      \\                        
LRc05\_E2\_3718&126&F8IV      &HD84737   &G0.5Va    &HD120136  &F6IV      \\                        
LRc09\_E2\_0548&126&F8IV      &HD10697   &G5IV      &HD128987  &G6V       \\                        
LRc03\_E2\_0935&130&G0V       &HD215648  &F7V       &HD204363  &F7 V      \\                        
LRc10\_E2\_5093&132&K0III     &HD170693  &K1.5III   &HD5286    &K1 IV     \\                        
LRc10\_E2\_0740&134&K3III     &HD157681  &K5III     &HD34255   &K4Iab:    \\                        
LRc10\_E2\_1984&141&       ---&       ---&       ---&HD126327  &M7.5 III  \\                        
LRc09\_E2\_0308&141&B1V       &HD157214  &G0V       &HD115617  &G5V       \\                        
LRc07\_E2\_0158&179&F8IV      &HD216385  &F7IV      &HD136064  &F9IV      \\                        
LRc08\_E2\_0275&179&G8V       &HD192699  &G5        &HD76813   &G9III     \\                        
SRc01\_E1\_0346&187&A0V       &HD41692   &B5IV      &HD168199  &B5 V      \\                        
LRc07\_E2\_0534&196&G5III     &HD180610  &K2III     &HD112127  &K2.5III   \\                        
LRc09\_E2\_0131&219&G0V       &HD168723  &K0III-IV  &HD149661  &K2V       \\                        
LRc07\_E2\_0307&226&F5IV      &HD173667  &F6V       &HD150012  &F5IV      \\                        
LRc07\_E2\_0146&240&F0V       &HD210855  &F8V       &HD136064  &F9IV      \\                        
LRc07\_E2\_0182&248&A5V       &HD120136  &F6IV      &HD61064   &F6III     \\                        
LRc06\_E2\_0119&265&F8IV      &HD187013  &F7V       &HD59380   &F8V       \\                        
LRc07\_E2\_0482&277&A5IV      &HD215648  &F7V       &HD62301   &F8V       \\                        
LRc07\_E2\_0187&293&F8IV      &HD215648  &F7V       &HD59380   &F8 V      \\                        
LRc05\_E2\_0168&322&F8IV      &HD176377  &G0        &HD39587   &G0V       \\                        
LRc05\_E2\_0168&437&F8IV      &HD176377  &G0        &HD39587   &G0V       \\                        

\hline
\end{tabular}
\end{table*}

The results of the classification are listed in Table~\ref{tab:classinfo}. First of all, there are some interesting qualitative findings. In neither case, the best-matching external template originates from the same star as the best-matching internal template, in spite of a high number of stars common to both libraries. Also it occurs that an external template spectrum matches another {\corot} object. The CFLIB spectrum of HD\,120136 matches LRc05\_E2\_3718 while the Nasmyth spectrum of HD\,120136 matches the spectrum of LRc07\_E2\_0182. Apparently, the CFLIB spectrum of HD\,120136 is as different from the Nasmyth spectrum as is the difference between the spectra of LRc07\_E2\_0182 and LRc05\_E2\_3718. Although both stars are late-F or early-G type stars, we note, that the SNR of the spectrum of LRc05\_E2\_3718 is only half of that of LRc07\_E2\_0182. Interestingly, the Nasmyth spectrum of HD\,84737 matched Nasmyth spectra of as many as five {\corot} objects. Again, the SNR is obviously important since all the five target spectra suffer from low signal!

\begin{figure}[h] 
\centering
\subfigure[\label{fig:sn_spec_nas}external templates (CFLIB) - internal templates]{\includegraphics[width=7cm]{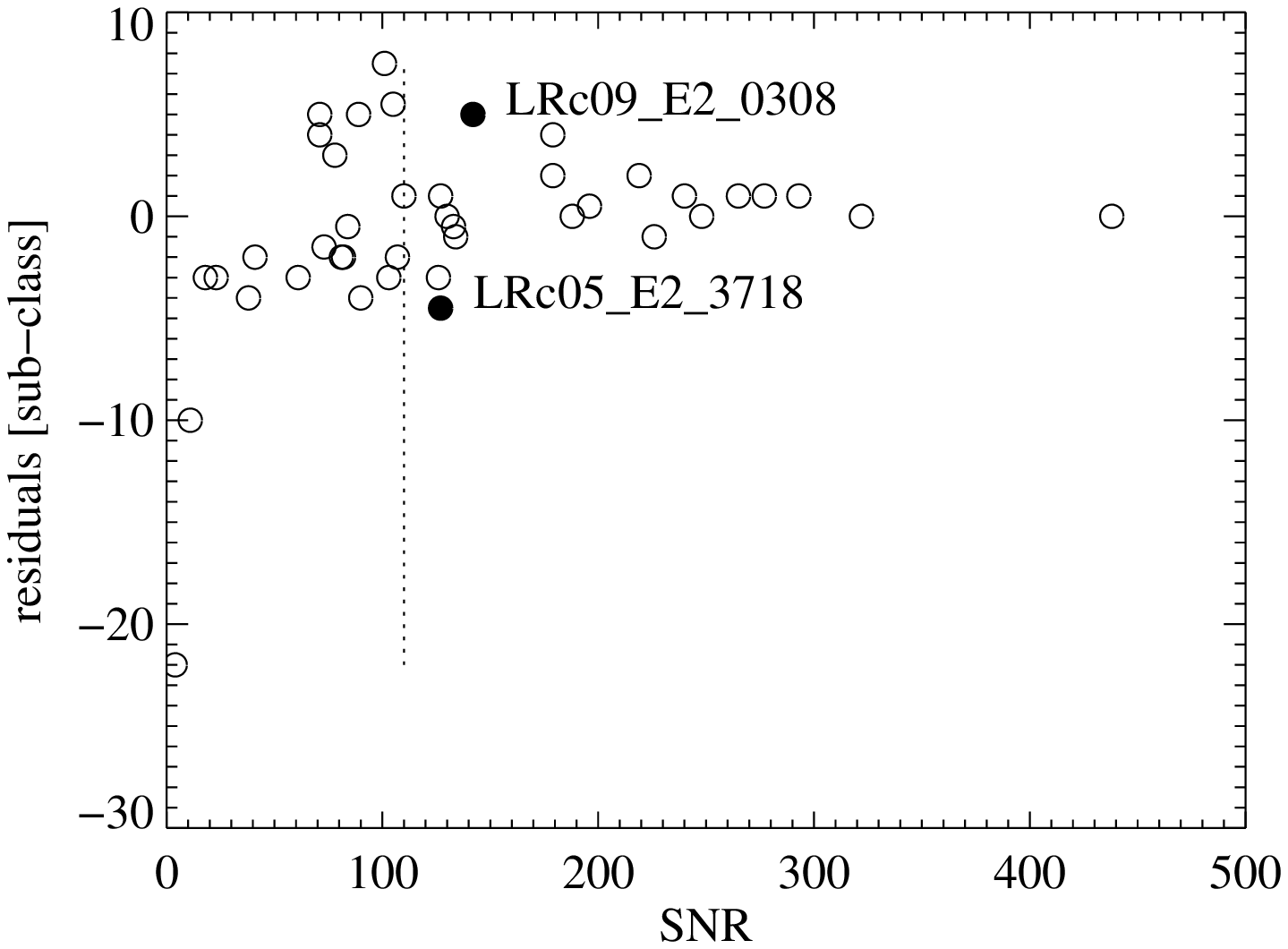}}
\subfigure[\label{fig:sn_spec_exo}photometric classification  ({\exodat}) - internal templates]{\includegraphics[width=7cm]{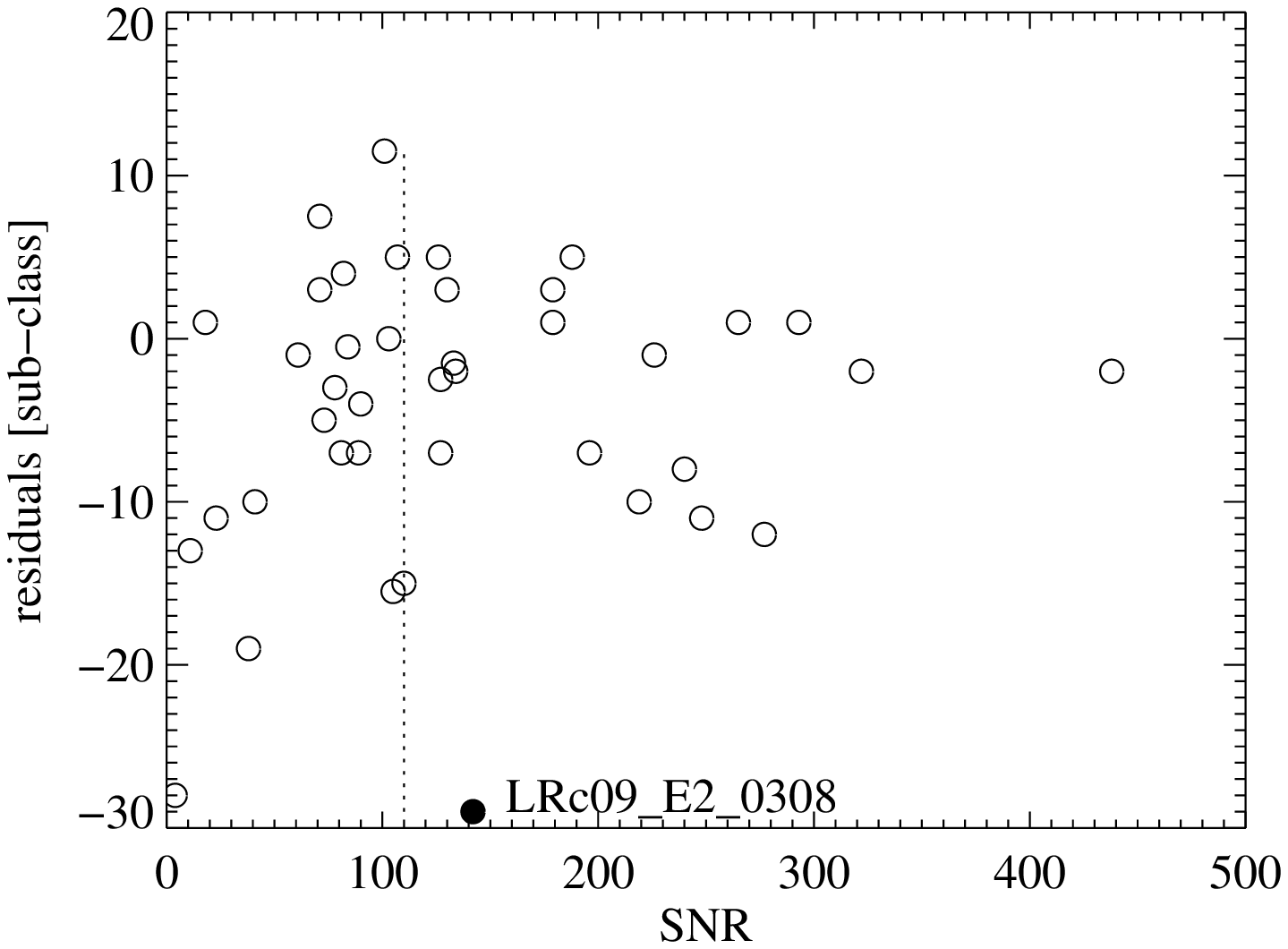}}
\subfigure[\label{fig:sn_spec_val}photometric classification ({\exodat}) - external templates (CFLIB)]{\includegraphics[width=7cm]{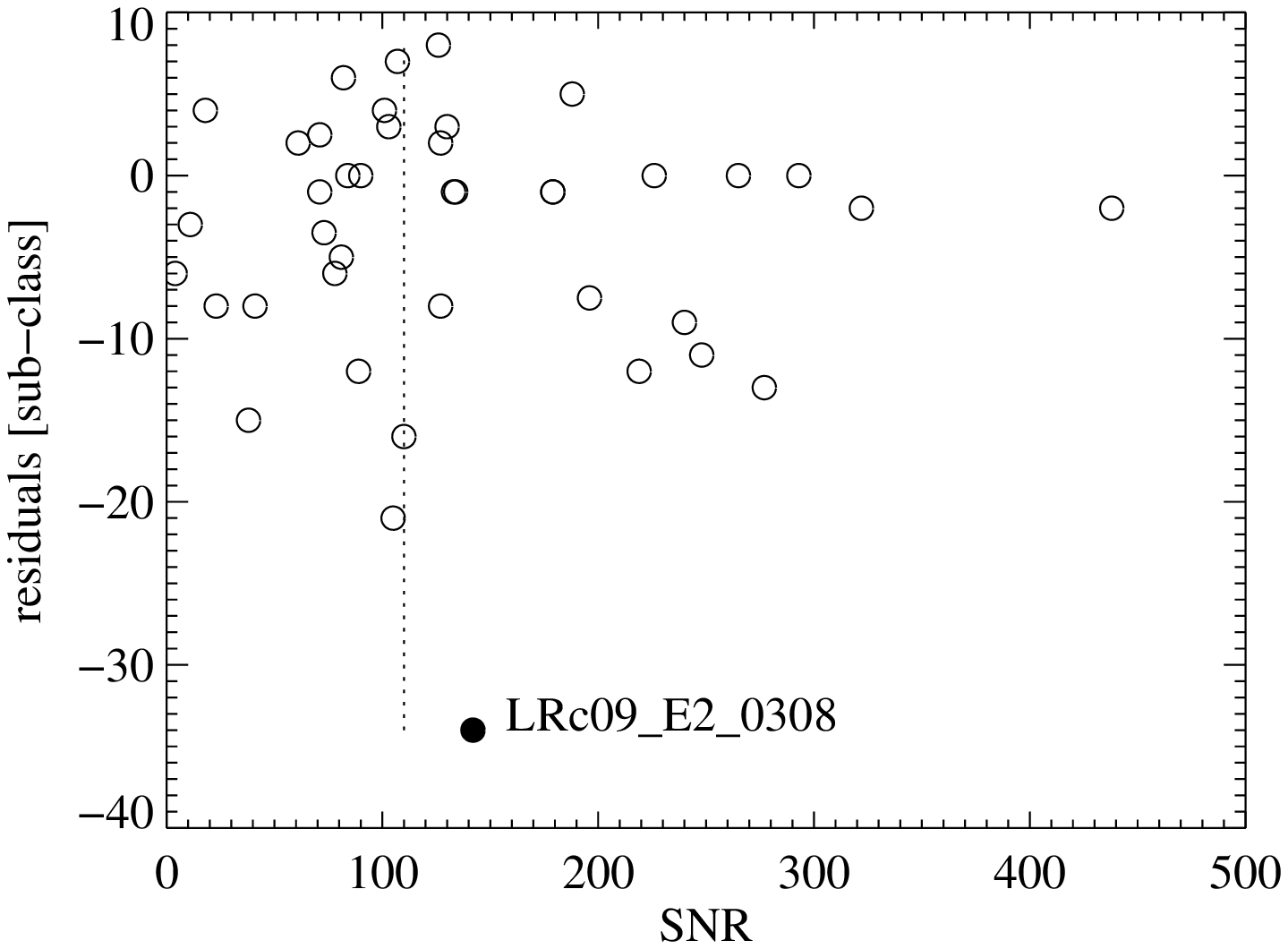}}
\caption{\label{fig:sn_spec} Residuals of classifications when using external templates and photometric classifications. The difference of spectral types (in sub-classes) is displayed vs. the SNR. The vertical line shows the median SNR found in Sect.~\ref{sect:obs}, and is used to distinguish good and bad spectra. The $2\sigma$ outliers among the good spectra are highlighted by filled circles and discussed in the text.}
\end{figure}

\begin{figure}[h] 
\centering
\includegraphics[width=7cm]{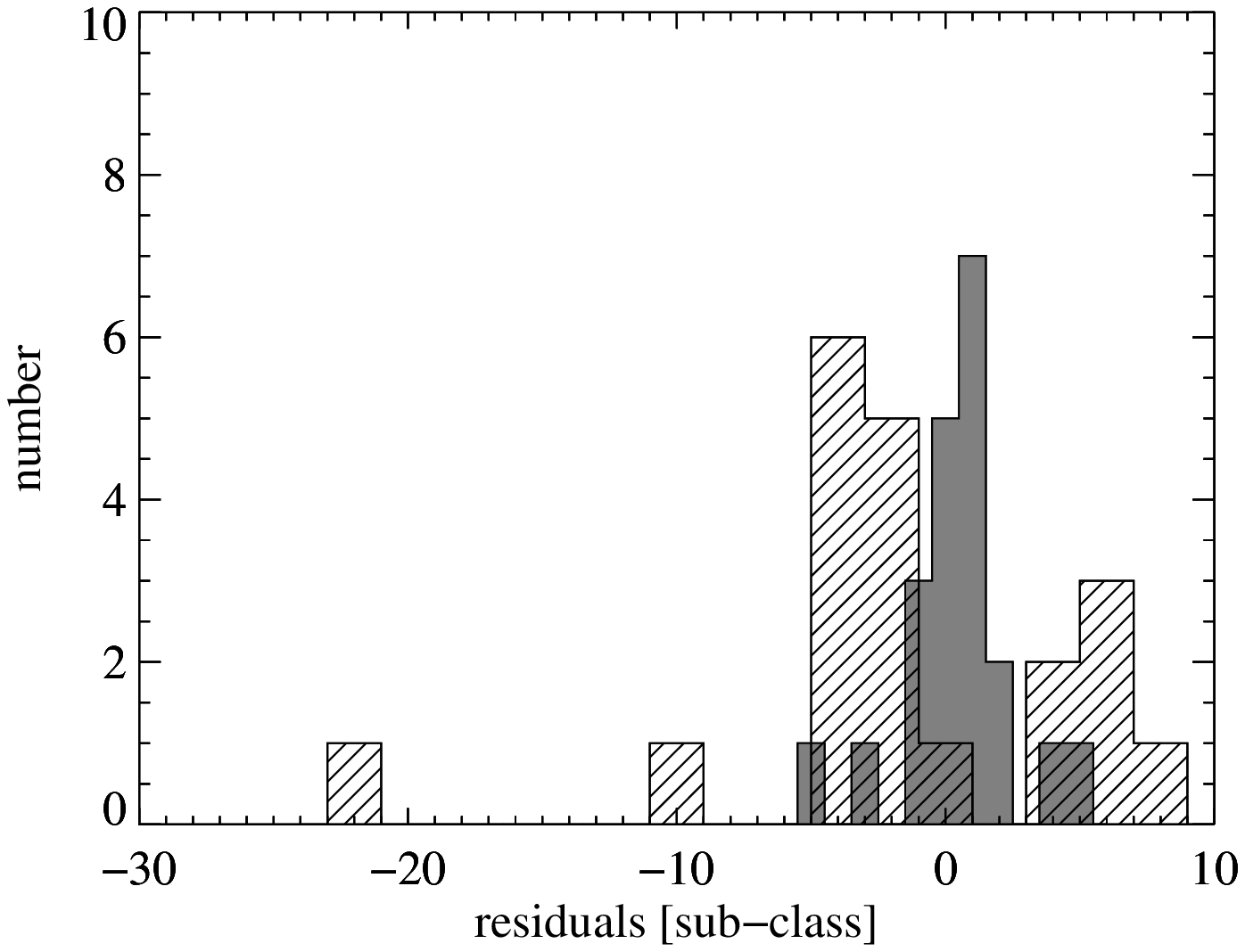}
\caption{\label{fig:hist_spec} Histogram of residuals of classifications when using the CFLIB library (corresponding to Fig.~\ref{fig:sn_spec_nas}). The difference of spectral types is shown in units of sub-class. The filled style distinguishes spectra with good (grey) and bad signal (hatched) according to the division line shown in Fig.~\ref{fig:sn_spec}.}
\end{figure}

The qualitative discussion above shows that the SNR plays a role. A more quantitative presentation of the data confirms that the agreement depends on the noise level of the {\corot} target spectra (Fig.~\ref{fig:sn_spec_nas}). Fig.~\ref{fig:hist_spec} presents another view on the data by projecting along the noise axis and showing the distribution of the residuals separately for the good and bad spectra. The spread tends to increase with decreasing signal, i.e. the strongest discrepancies are usually encountered for spectra of least quality. At highest signal, the discrepancies tend to vanish. 

The mean difference in spectral type is as low as 2 sub-classes when considering the good spectra only. Remarkably, hardly any systematic offsets or trends are seen in our sample.

The distribution of residuals appears far from normal in the case of the bad spectra. In the case of the good spectra, however, an identification of outliers seems possible. There are two outliers in spectral type, LRc09\_E2\_0308 and LRc05\_E2\_3718. The deviation cannot be readily explained since the SNR is reasonably good.

The spectroscopic classifications were compared with photometric classifications taken from the online version of {\exodat} as of May 16$^\mathrm{th}$, 2014. We note that the scatter is increased for bad spectra (Figs.~\ref{fig:sn_spec_exo}, \ref{fig:sn_spec_val}). But even for good spectra, the scatter is dramatically larger and we note that photometry tends to assign earlier spectral types. The level of agreement is the same whatever the choice of the template library - internal or external.

\begin{table}
\caption{\label{tab:lcnas} Contingency table showing the agreement of luminosity classifications obtained with the external CFLIB template library and the internal Nasmyth library. The number of classifications based on good spectra is shown and succeeded in brackets by the number of classifications based on bad spectra. Good and bad spectra are distinguished using the median of the SNR (cf. Fig.~\ref{fig:sn_spec}). The solid lines in the table divide giant-like classes and dwarf-like classes.}
\centering
\begin{tabular}{l|rrrr|rr}
\hline
&\multicolumn{6}{c}{Nasmyth templates}\\
CFLIB templates&I&II&III&III-IV&IV&V\\
\hline
I&&&1&&(1)&\\
II&&&(1)&&&(1)\\
III&&&1(3)&&1&1(2)\\
III-IV&&&&&&\\
\hline
IV&&&1&&1&4(2)\\
V&&(2)&&1&2&8(8)\\
\hline
\end{tabular}
\end{table}

\begin{table}
\caption{\label{tab:lcexo} Contingency table showing the agreement of spectroscopic and photometric luminosity classifications: internal Nasmyth catalogue vs. photometric Exodat classification. The layout follows Table~\ref{tab:lcnas}.}
\centering
\begin{tabular}{l|rrrr|rr}
\hline
&\multicolumn{6}{c}{Nasmyth templates}\\
Exodat&I&II&III&III-IV&IV&V\\
\hline
I&&&&&(1)&\\
II&&&(1)&&&\\
III&&&3(2)&&&(2)\\
III-IV&&&&&&\\
\hline
IV&&&&&2&8(6)\\
V&&(2)&(1)&1&2&5(5)\\
\hline
\end{tabular}
\end{table}

\begin{table}
\caption{\label{tab:lcval} Contingency table showing the agreement of spectroscopic and photometric luminosity classifications: CFLIB catalogue vs. photometric Exodat classification. The layout follows Table~\ref{tab:lcnas}.}
\centering
\begin{tabular}{l|rrrr|rr}
\hline
&\multicolumn{6}{c}{CFLIB templates}\\
Exodat&I&II&III&III-IV&IV&V\\
\hline
I&(1)&&&&&\\
II&&(1)&&&&\\
III&1&&1(2)&&1(2)&\\
III-IV&&&&&&\\
\hline
IV&&&(1)&&3&7(5)\\
V&&(1)&2(2)&&2&4(5)\\
\hline
\end{tabular}
\end{table}

\citet{sebastian12} showed that for solar-type stars the computer-based spectral classification cannot distinguish between main-sequence stars and sub-giants but between main-sequence stars and giants or super-giants. Therefore, we distinguish giant-like (I, II, III, III-IV) and dwarf-like luminosity classes (IV, V); also because of the low numbers in the present work. The luminosity classification based on the two different sets of templates is compared in a statistical sense (Table~\ref{tab:lcnas}). Each field contains the number of {\corot} targets with corresponding classifications. Again, good and bad spectra are distinguished. The total number of good(bad) spectra is 21(20). 15(10) stars are dwarf-like according to either set of templates while there is agreement on a giant-like luminosity class for 2(4) stars. Still, there is substantial disagreement for 4(6) stars. Two(four) dwarf-like stars are given giant-like classifications when using the CFLIB templates and 2(2) giant-like stars are assigned dwarf-like classifications. In summary, only 19\,\% of the classifications of good spectra are not reproduced by the external library compared to 30\,\% of the classifications of the bad spectra. 

The contingency tables are reproduced for the comparisons with the photometric classification from {\exodat} (Tables \ref{tab:lcexo}, \ref{tab:lcval}). When comparing the internal library with {\exodat}, there are 1(6) off-diagonal elements. Comparing the CFLIB to {\exodat}, there are 3(6) off-diagonal elements.

For an update of the status of {\corot} candidates with a spectroscopic classification, we recommend a rather conservative approach in order to not discard planet candidates prematurely. For the present sample and for the discussion below, we adopt a giant-like luminosity class only in the case that both template libraries yield a giant-like classification. Five out of these 6 stars were classified giants by photometry, too. Furthermore, we identified seven early-type stars (F3 or earlier), six out of these also early-type according to {\exodat}. 

However, there is disagreement in several cases. Intriguingly, seven stars are early-type according to photometry but late-type according to spectroscopy, affecting almost half of the photometric early-type targets. Furthermore, 2 out of 9 giants are actually dwarf stars. There is no single late-type star which turned out early-type. Only one early-type dwarf turned out a giant.

\section{Discussion}
\label{sect:disc}


We obtained target and template spectra with the same instrument in order to avoid any systematic effects which might be introduced by the use of another instrument. This involves a major effort since a large number of spectral types has to be covered. As many as 149 template stars have been observed to cover different luminosity classes and chemical abundance patterns. Although the coverage is not yet complete, it is very dense.

One of the main goals was to find out whether an external template catalogue, here the CFLIB, can be used to reproduce the results of a classification with an internal library. The best-matching template stars found in the present work show that the outcome mainly depends on the SNR of the target spectra. At sufficiently high SNR ($\ge100$), the mean difference in spectral type is still within the internal uncertainties of the method of a few sub-classes \citep{Guenther+2012,sebastian12}. Also a discrepancy of the luminosity classification occurred more often in the case of bad target spectra. Although this assessment is based on small numbers, too, it ascribes at least part of the discrepancy to noise.

The agreement in spectral class for good target spectra gives high confidence in the use of external templates. In contrast, the photometric classification deviates by up to an entire spectral class. Nevertheless, as the scatter w.r.t. photometry increases for spectra with bad SNR, we conclude - as is expected - that the spectral classification is not able to provide accurate spectral types if the signal is too low. The photometric {\exodat} classification of luminosity performs as well as a spectral classifiation employing the external library. In this respect, the present work extends the work of \citet{sebastian12} who derived spectral types from AAO spectra using templates from CFLIB and who compared the results to {\exodat} classifications.

The comparison of the spectroscopic and photometric classifications shows that photometry favours early-type classifications (cf. Fig.~\ref{fig:sn_spec}). Half of them turned out late-type stars which is roughly in line with the findings of \citet{sebastian12} who showed that 30\,\% of the photometric A and B stars are actually late-type stars. Planet search campaigns often remove early-type targets from the list of candidates since the radial velocity follow-up of such objects can be challenging. Moreover, the low-resolution spectroscopy was able to identify 2 dwarf stars among 9 photometric giants. Giant stars are excluded from follow-up since the transit signature would be due to a low-mass star rather than a planet. This way, low-resolution spectroscopy can recover good candidates among early-type targets (50\,\%) and giant stars (25\,\%) discarded otherwise. For the present work, we studied 17 late-type dwarf stars which are particularly interesting for follow-up. None of these turned out a giant so that photometry performs much better in this case.

\section{Summary and Conclusions}
\label{sect:concl}

Within the ground-based follow-up of {\corot} targets we took low-resolution spectra of 42 objects with the low-resolution Nasmyth spectrograph in Tautenburg. 

The spectra of the {\corot} targets were classified using two different sets of templates, an internal template library taken with the same instrument as the target spectra and the external CFLIB library. Although the internal library comprises spectra of 149 stars, the coverage is not fully complete and will be complemented in upcoming observing runs. The use of the new set of templates to refine preliminary classifications is intriguing. The new template grid is densest in the regime of F and G dwarfs which is most promising for planet detections with {\corot}. 

We found that the use of external library spectra yields similar results when the signal-to-noise ratio of the target spectrum is sufficiently high ($\ge\,100$). However, the luminosity classification with an external library does not perform better than a photometric classification. Therefore, it seems feasible to resort to a less costly photometric luminosity classification when an internal template library is not available.

As an aside, we know the atmospheric parameters of the matching templates. Therefore, a quantitative classification by atmospheric parameters becomes feasible, including effective temperature and surface gravity along with chemical abundances. This might disentangle the well-known degeneracy in spectral classification due to unknown metallicity that may certainly affect the present work. Also, the follow-up of transit-planet candidates would benefit a lot from quantitative classification. However, there are major uncertainties. At first, systematic tests will be needed to assess the accuracy of the quantitative spectral classification, e.g. by reproducing the atmospheric parameters of stars selected from the template library.

The present work highlights the importance of spectroscopic follow-up of {\corot} candidates when only a photometric classification has been available before. With modern multi-object spectrographs at hand (e.g. AAO, LAMOST), low-resolution spectroscopic characterisation should become an indispensable part of any effort to follow-up on planet candidates identified in wide-field surveys.

\acknowledgements
The selection of {\corot} targets is based on contributions by P. Bord\'e, F. Bouchy, R. Diaz, S. Grziwa, G. Montagnier, and B. Samuel within the detection and ground-based follow-up of CoRoT candidates. M.A. was supported by DLR (Deutsches Zentrum f\"ur Luft-und Raumfahrt) under the project 50 OW 0204. We would like to thank the workshops and the night assistants at the observatory in Tautenburg, Germany. This research has made use of the ExoDat Database, operated at LAM-OAMP, Marseille, France, on behalf of the CoRoT/Exoplanet program. This research has made use of the SIMBAD database, operated at the CDS, Strasbourg, France, and NASA’s Astrophysics Data System Bibliographic Services.

\end{document}